\documentclass[twocolumn,english,aps,prl]{revtex4}
\usepackage[T1]{fontenc}
\usepackage[latin9]{inputenc}
\usepackage{color}
\definecolor{document_fontcolor}{rgb}{0, 0, 0}
\color{document_fontcolor}
\usepackage{babel}
\usepackage{array}
\usepackage{bm}
\usepackage{multirow}
\usepackage{amsmath}
\usepackage{graphicx}
\usepackage[unicode=true,pdfusetitle,
 bookmarks=true,bookmarksnumbered=false,bookmarksopen=false,
 breaklinks=false,pdfborder={0 0 1},backref=false,colorlinks=true]
 {hyperref}

\makeatletter

\providecommand{\tabularnewline}{\\}

\@ifundefined{textcolor}{}
{%
 \definecolor{BLACK}{gray}{0}
 \definecolor{WHITE}{gray}{1}
 \definecolor{RED}{rgb}{1,0,0}
 \definecolor{GREEN}{rgb}{0,1,0}
 \definecolor{BLUE}{rgb}{0,0,1}
 \definecolor{CYAN}{cmyk}{1,0,0,0}
 \definecolor{MAGENTA}{cmyk}{0,1,0,0}
 \definecolor{YELLOW}{cmyk}{0,0,1,0}
}

\usepackage{babel}

\makeatother

\begin{document}

\title{Magnetization nutation induced by surface effects in nanomagnets}

\author{R. Bastardis}
\email{roland.bastardis@univ-perp.fr}

\selectlanguage{english}%

\author{F. Vernay}
\email{francois.vernay@univ-perp.fr}

\selectlanguage{english}%

\author{H. Kachkachi}
\email{hamid.kachkachi@univ-perp.fr}

\selectlanguage{english}%

\address{Laboratoire PROMES CNRS (UPR-8521) \& Université de Perpignan Via
Domitia, Rambla de la thermodynamique, Tecnosud, 66100 Perpignan,
France}

\date{\today}
\begin{abstract}
We investigate the magnetization dynamics of ferromagnetic nanoparticles
in the atomistic approach taking account of surface anisotropy and
the spin misalignment it causes. We demonstrate that such inhomogeneous
spin configurations induce nutation in the dynamics of the particle's
magnetization. More precisely, in addition to the ordinary precessional
motion with frequency $f_{p}\sim10\,{\rm GHz}$, we find that the
dynamics of the net magnetic moment exhibits two more resonance peaks
with frequencies $f_{c}$ and $f_{n}$ which are higher than the frequency
$f_{p}$: $f_{c}=4\times f_{p}\sim40\,{\rm GHz}$ is related with
the oscillations of the particle's magnetic moment between the minima
of the effective potential induced by weak surface anisotropy. On
the other hand, the much higher frequency $f_{n}\sim1\,{\rm THz}$
is attributed to the magnetization fluctuations at the atomic level
driven by exchange interaction. We have compared our results on nutation
induced by surface effects with those rendered by the macroscopic
approach based on the Landau-Lifshitz-Gilbert equation augmented by
an inertial term (proportional to the second-order time derivative
of the macroscopic moment) with a phenomenological coefficient. The
good agreement between the two models have allowed us to estimate
the latter coefficient in terms of the atomistic parameters such as
the surface anisotropy constant. We have thus proposed a new origin
for the magnetization nutations as being induced by surface effects
and have interpreted the corresponding resonance peaks and their frequencies.
\end{abstract}
\maketitle

\section{introduction}

Research on nanoscale magnetic materials benefits from a continuing
impetus owing to an increasing demand of our modern societies for
ever smaller devices with ever higher storage densities and faster
access times. These devices are the upshot of spintronics or magnonic
applications with materials exhibiting thermally stable magnetic properties,
energy efficient magnetization dynamics, and controlled fast magnetization
switching. In the macroscopic approach, the magnetization dynamics
on time scales ranging from microseconds to femtoseconds can be described
by the Landau-Lifshitz-Gilbert (LLG) equation \cite{lanlif35,gilbert56phd,gilbert04ieeetm}

\begin{equation}
\frac{d\bm{m}}{dt}=\bm{m}\times\left(\gamma\bm{H}_{{\rm eff}}-\frac{\alpha}{m}\frac{d\bm{m}}{dt}\right)\label{eq:LLGEqt}
\end{equation}
where $\bm{H}_{{\rm eff}}$ is the effective field acting on the macroscopic
magnetic moment $\bm{m}$ carried by the nanomagnet, $\gamma$ the
gyromagnetic factor and $\alpha$ the phenomenological damping parameter.
Equation (\ref{eq:LLGEqt}) describes the relaxation of $\bm{m}$
towards $\bm{H}_{{\rm eff}}$ while maintaining a constant magnitude,
\emph{i.e. }$\left\Vert \bm{m}\right\Vert =m$, assuming that the
nanomagnet is not coupled to any heat bath or other time-dependent
external perturbation. The first term on the right hand of Eq. (\ref{eq:LLGEqt})
describes the precessional motion of the magnetic moment $\bm{m}$
around the effective field $\bm{H}_{{\rm eff}}$. This is well known
from the classical mechanics of a gyroscope. Indeed, if an external
force tilts the rotation axis of the gyroscope away from the direction
of the gravity field, the rotation axis no longer coincides with the
angular-momentum direction. The consequence is an additional movement
of the gyroscope around the axis of the angular momentum. This motion
is called \emph{nutation}. In the case of the magnetic moment $\bm{m}$,
this additional motion (nutation) can occur if the effective field
$\bm{H}_{{\rm eff}}$ becomes time-dependent. Indeed, in the presence
of a time-dependent magnetic field (rf or microwave field), there
appears the fundamental effect of transient nutations which has been
widely investigated in NMR \cite{Torrey49pr}, EPR \cite{verfes73jcp,atkinsEtal74cpl},
and optical resonance \cite{hoctan68prl}, see also the review by
Fedoruk \cite{fedoruk02jas}. Magnetic or spin nutation was first
predicted in Josephson junctions \cite{zhufra06jpcm,franzhu08njp,franson08nanotech,PhysRevB.71.214520,PhysRevLett.92.107001}
and was later developed using various approaches based on first principles
\cite{mondaletal17prb}, electronic structure theory \cite{PhysRevLett.108.057204,PhysRevB.84.172403,PhysRevB.92.184410,thoningetal17sp,chengetal17prb},
or in a macrospin approach where the LLG equation (\ref{eq:LLGEqt})
is extended by a second-order time derivative \cite{ciorneietal11prb,oliveetal12apl,doi:10.1063/1.4921908,oliweg16jpcm}. 

Magnetic nutation may also occur at the level of atomic magnetic moments
on ultra-short time scales. For instance, in Ref. \onlinecite{bothen12prb}
it is argued that nutation is enhanced for atomic spins with low coordination
numbers and that it occurs on a time scale of the magnetic exchange
energy, \emph{i.e.} a few tens of femtoseconds. More generally this
spin nutation is caused by a nonuniform spin configuration which leads
to an inhomogeneous effective field $\bm{H}_{{\rm eff}}$ whose magnitude
and orientation are different for different lattice sites. These spatial
inhomogeneities are a typical result of surface effects that become
very acute in nanoscale magnetic systems such as magnetic nanoparticles.
In this work we adopt this atomistic approach and show that for a
magnetic nanoparticle regarded as a many-spin system, a model henceforth
referred to as the \emph{many-spin problem} (MSP), surface effects
do induce nutations of the net magnetic moment of the nanoparticle.
More precisely, this approach involves at least three energy scales,
namely the core (magneto-crystalline) anisotropy, the surface anisotropy
and exchange coupling. Consequently, there appear at least three different
frequencies: the lowest corresponds to the ordinary precession around
a fixed axis with a constant projection of the net magnetic moment
on the latter and the other two frequencies correspond to nutations
with a time-dependent projection of $\bm{m}$. In the limiting case
of weak surface effects, inasmuch as the spin configuration inside
of the nanomagnet can be regarded as quasi-collinear, the dynamics
of the nanomagnet can be described with the help of an effective macroscopic
model for the net magnetic moment of the nanomagnet. This model will
be referred to in the sequel as the \emph{effective one-spin problem}
(EOSP). More precisely, it has been shown that a many-spin nanomagnet
of a given lattice structure and energy parameters (on-site core and
surface anisotropy, local exchange interactions) can approximately
be modeled by a macroscopic magnetic moment $\bm{m}$ evolving in
an effective potential \cite{garkac03prl} that comprises second and
fourth powers of the components $m_{\alpha},\alpha=x,y,z$. Within
this approach we find two precession frequencies $f_{p}$ and $f_{c}$:
$f_{p}$ corresponds to the precession of $\bm{m}$ around the reference
$z$ axis with constant $m_{z}$ and $f_{c}$ to the frequency of
oscillations of $\bm{m}$ between the $4$ minima of the effective
potential produced by its quartic term. When surface or boundary effects
are too strong, the spin configuration can no longer be considered
as quasi-collinear and thereby the effective model is no longer a
good approximation, one has to take account of higher-order fluctuations
of the atomic spins. Doing so numerically, we find an additional nutation
frequency $f_{n}$ which is much higher than $f_{p}$ and $f_{c}$
as it corresponds to a movement of the atomic spins that occurs at
the time scale of the magnetic exchange interaction. 

Observation of nutation in magnetization dynamics is difficult because
the effect is rather small and the corresponding frequency is beyond
the detection capabilities of standard techniques using the magnetization
resonance such as the standard FMR or a network analyzer with varying
frequency. Nevertheless, from the high-frequency FMR ($115-345\,{\rm GHz}$)
spectra obtained for ultra-fine cobalt particles, the authors of Ref.
\onlinecite{respaudetal99prb} inferred low values for the transverse
relaxation time $\tau_{\perp}$ (two orders of magnitude smaller than
the bulk value) and suggested that this should be due to inhomogeneous
precession which possibly originates from surface spin disorder. Likewise,
in Ref. \onlinecite{bothen12prb} it was shown that nutation in magnetization
dynamics of nanostructures occurs at edges and corners, with a much
smaller amplitude than the usual precession. More recently, Li \emph{et
al.} \cite{PhysRevB.92.140413} performed HF-FMR measurements of the
effective magnetic field and showed that there was an additional contribution
which is quadratic in frequency as obtained from the additional term
$d^{2}\bm{m}/dt^{2}$ in the LLG equation \cite{PhysRevB.83.020410,Olive_JAP_2012}.

To sum up, in this work we first demonstrate that surface effects
or, more generally, non-collinear atomic spin ordering induce nutation
in the magnetization dynamics of a nanomagnet. Second, it establishes
a clear connection between nutation within our atomistic approach
and that described by the quadratic frequency dependence of the effective
field as described within the macroscopic approach including magnetization
inertia. If we cannot provide an analytical connection between the
corresponding parameters, we do provide a numerical correspondence
between the phenomenological parameter of the macroscopic approach
and our atomistic parameters such as the surface anisotropy constant.
We also propose an intermediate macroscopic model which accounts for
all three resonance frequencies. Finally, we speculate that the resonance
peak at $f_{c}$, induced by surface effects, provides a route for
observing nutation in well prepared assemblies of nanomagnets. 

The paper is organized as follows: in Section \ref{sec:Model} we
present our model of many-spin nanomagnets, discuss the effects of
surface anisotropy on the magnetization dynamics, and present our
main results showing two new resonance peaks which we attribute to
two kinds of magnetization nutation. In Subsection \ref{subsec:MSP}
we also discuss a particular situation where it is possible to analytically
derive the equation of motion of the net magnetic moment of the (many-spin)
nanomagnet which makes it clear that nutation is related with the
spin fluctuations at the atomic level. In Subsection \ref{subsec:IOSP}
we compare our results with other works in the literature mostly based
on the macroscopic approach using the Landau-Lifshitz-Gilbert equation
augmented by an inertial term, and establish a quantitative relationship
between the corresponding sets of parameters. Finally, in Section
\ref{sec:Conclusion-and-perspectives} we summarize the main results
of this work and then discuss the possibility to observe the magnetization
nutations in resonance experiments.

\section{Model and hypothesis\label{sec:Model}}

We consider a nanomagnet with $\mathcal{N}$ atomic spins $\bm{s}_{i}$
on a simple cubic lattice described by the (classical) Hamiltonian
($\left\Vert \bm{s}_{i}\right\Vert =1$)

\begin{equation}
\mathcal{H}=-\frac{1}{2}\sum_{i,j}J_{ij}\bm{s}_{i}\cdot\bm{s}_{j}-\bm{h}\cdot{\displaystyle {\displaystyle \sum_{i=1}^{\mathcal{N}}}}\bm{s}_{i}-{\displaystyle {\displaystyle \sum_{i=1}^{\mathcal{N}}}}\mathcal{H}_{{\rm an},i}\label{eq:Hamiltonian}
\end{equation}
where $\bm{h}=\mu_{a}\bm{H},$ $\mu_{a}$ is the magnetic moment associated
with the atomic spin, $\bm{H}$ is the magnetic field, $J_{ij}$ is
the exchange interaction (that may be different for core-surface,
surface-surface and core-core links), and $\mathcal{H}_{{\rm an},i}$
is the anisotropy energy at site $i,$ a function of $\bm{s}_{i}$
satisfying the symmetry of the problem. More precisely, $\mathcal{H}_{{\rm an},i}$
is the energy of on-site anisotropy which is here taken as uniaxial
for core spins and of Néel's type for surface spins \cite{Neel_anisotropy},
\emph{i.e.} 

\begin{equation}
\mathcal{H}_{{\rm an},i}=\left\{ \begin{array}{lc}
-K_{c}\left(\bm{s}_{i}\cdot\bm{e}_{z}\right)^{2}, & i\in{\rm core}\\
\\
+\frac{1}{2}K_{s}{\displaystyle \sum_{j\in{\rm n.n.}}}\left(\bm{s}_{i}\cdot\bm{e}_{ij}\right)^{2}, & i\in\mathrm{surface},
\end{array}\right.\label{eq:HamUA-NSA}
\end{equation}
where $\bm{e}_{ij}$ is the unit vector connecting the nearest neighbors
at sites $i$ and $j$ and $K_{{\rm c}}>0$ and $K_{{\rm s}}>0$ are
respectively the core and surface anisotropy constants.

The spin dynamics is described by the Landau-Lifshitz equation (LLE)
for the atomic spin $\bm{s}_{i}$
\begin{equation}
\frac{d\bm{s}_{i}}{d\tau}=\bm{s}_{i}\times\bm{h}_{{\rm eff},i}-\alpha\bm{s}_{i}\times\left(\bm{s}_{i}\times\bm{h}_{{\rm eff},i}\right),\label{eq:LLE-MSP}
\end{equation}
with the (normalized) local effective field $\bm{h}_{{\rm eff},i}$
acting on $\bm{s}_{i}$ being defined by $\bm{h}_{{\rm eff},i}=-\delta\mathcal{H}/\delta\bm{s}_{i}$;
$\tau$ is the reduced time defined by 
\begin{equation}
\tau\equiv\frac{t}{\tau_{{\rm s}}},\label{eq:ReducedTime}
\end{equation}
where $\tau_{{\rm s}}=\mu_{a}/\left(\gamma J\right)$ is a characteristic
time of the system's dynamics. By way of example, for cobalt $J=8{\rm \ meV}$
leading to $\tau_{s}=70\ {\rm fs}$. Henceforth, we will only use
the dimensionless time $\tau$. In these units, $\bm{h}_{{\rm eff},i}=\mu_{a}\bm{H}_{{\rm eff},i}/J$.

Equation (\ref{eq:LLE-MSP}) is a system of $2\mathcal{N}$ coupled
equations for the spins $\bm{s}_{i}$, $i=\left\{ 1,\cdots,\mathcal{N}\right\} $.
In this work, it is solved using iterative optimized second-order
methods using Heun's algorithm.

The particle's net magnetic moment is defined as 
\begin{equation}
\bm{s}_{0}=\frac{1}{\mathcal{N}}\sum_{i=1}^{\mathcal{N}}\bm{s}_{i}.\label{eq:Macrospin}
\end{equation}

Next, we introduce the verse of $\bm{s}_{0}$
\begin{equation}
\bm{m}\equiv\frac{1}{s_{0}}\bm{s}_{0},\qquad s_{0}=\left\Vert \frac{1}{\mathcal{N}}\sum_{i}\bm{s}_{i}\right\Vert .\label{eq:MacrospinVerse}
\end{equation}

As discussed in the introduction, because of surface effects or spatial
inhomogeneities of the effective field (mainly due to the fact that
the anisotropy constant and the easy axis depend on the lattice site),
the spin configuration is not uniform, for an arbitrary set of energy
parameters. As a consequence, the vectors $\bm{s}_{i}$ are not all
parallel to each other and as such we may define their deviation from
the direction $\bm{m}$ as \cite{garkac09prb}
\begin{align*}
\bm{s}_{i} & =\left(\bm{m}\cdot\bm{s}_{i}\right)\bm{m}+\bm{\psi}_{i}
\end{align*}
where we have introduced the vector
\[
\bm{\psi}_{i}\equiv\bm{s}_{i}-\left(\bm{m}\cdot\bm{s}_{i}\right)\bm{m}.
\]

It can be easily checked that $\bm{\psi}_{i}$ is perpendicular to
$\bm{s}_{i}$, \emph{i.e.} $\bm{\psi}_{i}\cdot\bm{s}_{i}=0=\bm{\psi}_{i}\cdot\bm{m}$
and satisfies $\sum_{i=1}^{\mathcal{N}}\bm{\psi}_{i}=\bm{0}$. This
means that the transverse vector $\bm{\psi}_{i}$ contains the Fourier
components with $\mathbf{k\neq0}$ and describes spin waves in the
nanomagnet. Whereas in the standard spin wave theory $\bm{s}_{0}$
is a constant corresponding to the ground-state orientation, here
it is treated as a time-dependent variable. 

Note that using the condition $\left\Vert \bm{s}_{i}\right\Vert =1$,
we may write $\bm{s}_{i}=\bm{m}\sqrt{1-\bm{\psi}_{i}^{2}}+\bm{\psi}_{i}$.
Now, in the realistic case, $K_{{\rm s}}\ll J$, the deviations of
$\bm{s}_{i}$ from the homogeneous state $\bm{m}$ are small and one
can adopt the following approximation
\[
\bm{s}_{i}\simeq\bm{m}\left(1-\frac{1}{2}\bm{\psi}_{i}^{2}\right)+\bm{\psi}_{i}\equiv\bm{m}+\delta\bm{s}_{i}
\]
where 
\begin{equation}
\delta\bm{s}_{i}\equiv-\frac{1}{2}\bm{\psi}_{i}^{2}\bm{m}+\bm{\psi}_{i}.\label{eq:SpinDeficit}
\end{equation}
Then, we define the magnetization deficit due to surface anisotropy
as follows

\begin{align}
\Delta m & \equiv-\frac{1}{\mathcal{N}}\sum_{i}\bm{m}\cdot\delta\bm{s}_{i}.\label{eq:MagDeficit}
\end{align}
Using Eq. (\ref{eq:SpinDeficit}) and $\sum_{i=1}^{\mathcal{N}}\bm{\psi}_{i}=\bm{0}$
we obtain 
\begin{equation}
\Delta m=\frac{1}{2\mathcal{N}}\sum_{i}\bm{\psi}_{i}^{2}=1-\frac{1}{\mathcal{N}}\sum_{i}\left(\bm{m}\cdot\bm{s}_{i}\right)=1-s_{0}.\label{eq:MagDeficit-v2}
\end{equation}

In what follows, we will show that the magnetization nutations are
a consequence of the magnetization deficit $\Delta m$ (which is due
to the transverse spin fluctuations $\bm{\psi}_{i}$) with respect
to $s_{0}$. In order to study nutation, we compute $\Delta m\left(\tau\right)$
or the components $m_{\alpha}\left(\tau\right)$, with $\alpha=x,y,z$.
In the sequel, we will mainly study the latter as their behavior clearly
illustrates the precession and nutation phenomena. In the next section
we present a sample of our results obtained for a cube-shaped nanomagnet
described by the Hamiltonian (\ref{eq:Hamiltonian}) together with
the anisotropy model in Eq. (\ref{eq:HamUA-NSA}).

\subsection{\label{subsec:MSP}Magnetization nutation induced by surface anisotropy}

In order to clearly illustrate the central result of this work, namely
that spin noncolinearities, induced by surface anisotropy, lead to
nutation in the magnetization dynamics of a nanomagnet, we consider
a simple shape, \emph{e.g.} a cube. Today, nanocubes (of iron or cobalt)
are routinely investigated in experiments since their synthesis has
become fairly well controlled \cite{snoecketal08nl,trunovaetal08jap,jiangetal10jac,mehdaouietal10j3m,kronastetal11nl,okellyetal12nanotech}.
Here we consider a nanocube of $\mathcal{N}=729$ spins located on
the vertices of a simple cubic lattice (\emph{i.e.} $N_{x}=N_{y}=N_{z}=9$).
This choice has the main advantage that the number of core spins ($\mathcal{N}_{c}=343$)
is comparable to that of surface spins ($\mathcal{N}_{s}=386$), a
configuration suitable for studying the role of surface effects versus
core properties. Then, we compute the time evolution of the net magnetic
moment $\bm{m}$ by solving the system of equations (\ref{eq:LLE-MSP}),
using Eqs. (\ref{eq:Macrospin}) and (\ref{eq:MacrospinVerse}). We
start from the initial state ${\bf s}_{i}\left(t=0\right)=\left(1/2,\ 1/2,\ 1/\sqrt{2}\right)$
which corresponds to all spins tilted to the same angle with respect
to the $z$ axis of the laboratory frame. 

Let us first consider the case of a nanocube with anisotropy energy
defined in Eq. (\ref{eq:HamUA-NSA}), \emph{i.e.} uniaxial for core
spins and of Néel's type for surface ones. A surface spin is defined
as the spin whose coordination number is smaller than in the core
(here $6$ on a simple cubic lattice). For simplicity, we set all
exchange couplings equal to a reference value $J$ everywhere in the
core, on the surface and at the interface between them, \emph{i.e.}
$J_{{\rm cc}}=J_{{\rm cs}}=J_{{\rm ss}}=J$. All energy constants
are then measured in units of $J$, so that $J=1$ and $k_{{\rm c}}\equiv K_{{\rm c}}/J=0.001$,
$k_{{\rm s}}\equiv K_{{\rm c}}/J=0.01$. These are typical values
extracted from experiments on cobalt and iron nanoparticles \citep{urquhartetal88jap,skocoe99iop,perrai05springer}.
In this calculation, the external magnetic field and damping are both
set to zero. 

\begin{figure}[t]
\begin{centering}
\includegraphics[width=0.95\columnwidth]{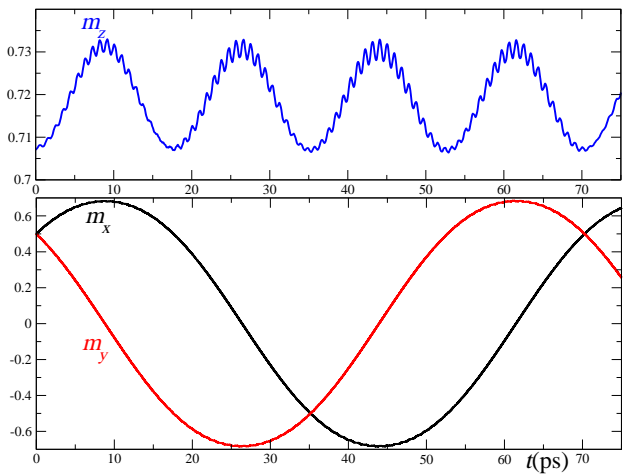} 
\par\end{centering}
\caption{\label{fig:MSP_Neel}Time evolution of the average magnetization components
$\left(m_{x},\ m_{y},\ m_{z}\right)$ for a nanomagnet of $\mathcal{N}=9\times9\times9=729$
spins with uniaxial anisotropy in the core and Néel surface anisotropy
on the surface. }
\end{figure}

\begin{figure}[t]
\begin{centering}
\includegraphics[width=0.5\columnwidth]{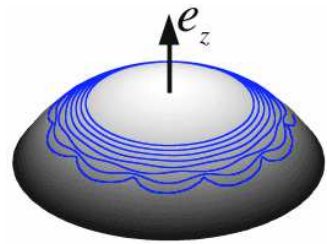} 
\par\end{centering}
\caption{\label{fig:DampedNutation}Illustration of the nutation of the macrospin
$\bm{s}_{0}$ in the presence of damping ($\alpha\protect\neq0$).
We have used nonzero damping for later reference.}
\end{figure}

Solving the LLE (\ref{eq:LLE-MSP}) renders the components of $\bm{m}\left(\tau\right)$
defined in Eq. (\ref{eq:MacrospinVerse}). These are shown in Fig.
\ref{fig:MSP_Neel}. In the lower panel, $m_{x}\left(\tau\right)$
and $m_{y}\left(\tau\right)$ show the usual precessional movement
of  $\bm{m}\left(\tau\right)$ around the $z$ axis. The corresponding
frequency for the parameters given above is $f_{p}=14\ {\rm GHz}$.
If $\bm{m}\left(\tau\right)$ were to exhibit only this precession,
its component $m_{z}\left(\tau\right)$ would be a constant with a
constant tilt angle between $\bm{m}\left(\tau\right)$ and the $z$
axis. However, as can be seen in the upper panel, it is clearly not
the case. Indeed, we see a double modulation of $m_{z}\left(\tau\right)$
in time; there are two oscillations: i) one with frequency $f_{c}=4\times f_{p}=56\ {\rm GHz}$
and an amplitude that is an order of magnitude smaller than precession,
and ii) another oscillation with the much higher frequency $f_{n}=1.1\ {\rm THz}$
and an amplitude two orders of magnitude smaller than precession.
These oscillations are further illustrated in Fig. \ref{fig:DampedNutation}.

Let us now discuss the origin of these oscillations. As discussed
in the introduction, in the case of not-too-strong surface effects,
the MSP may be mapped onto an EOSP \cite{garkac03prl,kacbon06prb,yanesetal07prb,kachkachi07j3m}
for the net magnetic moment $\bm{m}$ of the particle evolving in
an effective potential containing a quadratic and a quartic term in
the components of $\bm{m}$. This work has recently been extended
to cube-shaped magnets \cite{Garanin_arxiv_2018}. So for a nanomagnet
within the EOSP approach the equation of motion reads 

\begin{equation}
\frac{d\bm{m}}{d\tau}=\bm{m}\times\left[2k_{2}m_{z}\bm{e}_{z}-4k_{4}\left(m_{z}^{3}\bm{e}_{z}+m_{y}^{3}\bm{e}_{y}+m_{x}^{3}\bm{e}_{x}\right)\right].\label{eq:LLE-EOSP}
\end{equation}

Here $z=6$ is the coordination number and $k_{2}=k_{c}\mathcal{N}_{c}/\mathcal{N}$.
For a sphere $k_{4}=\kappa k_{s}^{2}/zJ$ where $\kappa$ is a surface
integral \cite{garkac03prl}, and for a cube we have $k_{4}=\left(1-0.7/\mathcal{N}^{1/3}\right)^{4}k_{s}^{2}/zJ$
\cite{Garanin_arxiv_2018}.

The components $m_{\alpha}\left(\tau\right)$ rendered by Eq. (\ref{eq:LLE-EOSP})
exhibit two resonance peaks corresponding to: i) the ordinary precession
with frequency $f_{p}$ and ii) the oscillation with frequency $f_{c}$
between the minima of the effective potential induced by the term
in $k_{4}$. The latter is due to the fact that the effective magnetic
moment has now to explore a potential-energy surface that comprises
four saddle points because of the cubic anisotropy (with constant
$k_{4}$). Therefore, $m_{z}$ visits a minimum each time $\bm{m}$
passes over one of these saddle points, and this occurs with the frequency
$f_{c}=4\times f_{p}=56\ {\rm GHz}$. Thus, $f_{c}$ is a consequence
of the first correction stemming from (relatively weak) surface effects. 

\begin{figure}[t]
\begin{centering}
\includegraphics[scale=0.3]{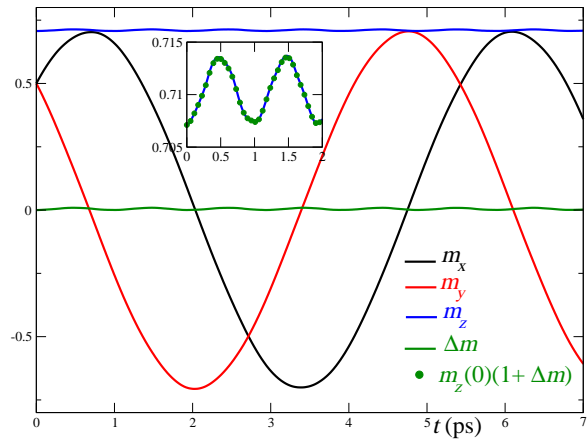} 
\par\end{centering}
\caption{\label{fig:Fig_m0z_Dm}Time evolution of the net magnetic moment compared
with that of the magnetization deficit. The exchange parameters are
homogeneous ($J=J_{cs}=J_{s}=1$), both surface and core spins have
a uniaxial anisotropy along the $z$ axis with surface anisotropy
$k_{s}=0.1$ and core anisotropy $k_{c}=0.01$.}
\end{figure}

In the case of larger values of $k_{s}$ and thereby stronger spin
noncolinearities, it is no longer possible to map the many-spin particle
onto an effective particle. One then has to fully deal with the spin
fluctuations. As a consequence it is no longer an easy matter to derive
an equation of motion similar to Eq. (\ref{eq:LLE-EOSP}) in the general
case. Nevertheless, in Ref. \onlinecite{garkac09prb} two relatively
simpler configurations of anisotropy were studied, namely a uniform
uniaxial anisotropy (with the same constant and orientation) or a
random anisotropy (with the same constant and random orientation).
It was then possible to derive a system of (coupled) equations for
$\bm{m}\left(t\right)$ and $\bm{\psi}_{i}\left(t\right)$ containing
higher-order terms in $\bm{\psi}_{i}\left(t\right)$, see Eqs. (21)
and (26) in Ref. \onlinecite{garkac09prb}. In the present situation
with a nonuniform anisotropy configuration, these higher-order contributions
are responsible for the nutation movement with frequency $f_{n}$,
as they lead to a net magnetization deficit, see Eq. (\ref{eq:MagDeficit-v2})
and Fig. \ref{fig:Fig_m0z_Dm} where the plot of $\Delta m$ shows
such a movement. More precisely, these fluctuations of the atomic
spins lead to a precession of the latter around their local effective
field $\bm{h}_{{\rm eff},i}$ that evolves in time due to exchange
interaction. Unfortunately, in this complex situation it is a rather
difficult task to derive an explicit expression for $\bm{h}_{{\rm eff},i}$
and thereby an analog of Eq. (\ref{eq:LLE-EOSP}). However, we may
consider a simpler model of a nanomagnet with a uniaxial anisotropy
having the easy axis along $\bm{e}_{z}$ for all sites, but with a
constant that is different in the core from that on the surface, \emph{i.e.}
$\bm{e}_{i}\parallel\bm{e}_{z},\,k_{{\rm c}}\neq k_{{\rm s}}$. Therefore,
instead of the model in Eq. (\ref{eq:HamUA-NSA}) we consider the
following one

\begin{equation}
\mathcal{H}_{{\rm an},i}=\left\{ \begin{array}{lc}
-k_{c}\left(\bm{s}_{i}\cdot\bm{e}_{z}\right)^{2}, & i\in{\rm core},\\
\\
-k_{s}\left(\bm{s}_{i}\cdot\bm{e}_{z}\right)^{2}, & i\in\mathrm{surface}.
\end{array}\right.\label{eq:HamUA-UA}
\end{equation}

This configuration is quite plausible especially in elongated nanomagnets
such as nanorods \cite{cordenteetal01nanolett} and nanowires \cite{camaraetal16apl}
where the magnetostatic energy is strong enough to induce an effective
uniaxial anisotropy along the major axis of the nanomagnet.

Then, it is possible to derive a system of equations for $\bm{m}$
and $\bm{\psi}_{i}$ (to second order in $\bm{\psi}_{i}$). The equation
for $\bm{\psi}_{i}$ is cumbersome and thus omitted here as it is
not necessary to the discussion that follows. That of $\bm{m}$ reads
\begin{eqnarray}
\frac{d\bm{m}}{d\tau} & \simeq & \bm{m}\times\frac{2}{\mathcal{N}}\sum_{i}k_{i}\left(m_{z}+\psi_{z,i}-m_{z}\psi_{i}^{2}\right)\bm{e}_{z}\nonumber \\
 & + & \bm{m}\times\frac{2}{\mathcal{N}}\sum_{i}k_{i}\left[\frac{1}{\mathcal{N}}\sum_{j}\left(m_{z}\frac{\psi_{j}^{2}}{2}\right)\right]\bm{e}_{z}\nonumber \\
 & - & \bm{m}\times\frac{2}{\mathcal{N}}\sum_{i}k_{i}\left[\left(m_{z}\right)^{2}+m_{z}\psi_{z,i}\right]\bm{\psi}_{i}.\label{eq:m0_UA}
\end{eqnarray}

First, setting $\bm{\psi}_{i}=\bm{0}$ above we obtain $d\bm{m}/d\tau=\bm{m}\times\frac{1}{\mathcal{N}}\sum_{i}\left(2k_{i}\right)m_{z}\bm{e}_{z}=\bm{m}\times2k_{{\rm eff}}m_{z}\bm{e}_{z}$,
which describes the precession of $\bm{m}$ around the effective field
$\bm{h}_{{\rm eff}}$ with 
\begin{equation}
\bm{h}_{{\rm eff}}=2k_{{\rm eff}}m_{z}\bm{e}_{z},\quad k_{{\rm eff}}=\frac{\mathcal{N}_{c}k_{c}+\mathcal{N}_{s}k_{s}}{\mathcal{N}}.\label{eq:effective_parameters-1}
\end{equation}
This clearly shows that nutation disappears in the absence of the
spin fluctuations $\bm{\psi}_{i}$. Furthermore, projection on the
$z$ axis of Eq. (\ref{eq:m0_UA}) yields the relation $dm_{z}/d\tau\simeq m_{z}d\left(\Delta m\right)/d\tau$,
where $\Delta m$ is the magnetization deficit defined in Eqs. (\ref{eq:MagDeficit},
\ref{eq:MagDeficit-v2}). Upon integrating over time we obtain (to
lowest order in $\bm{\psi}_{i}$)
\begin{eqnarray}
m_{z}\left(\tau\right) & \simeq & m_{z}\left(0\right)\left[1+\Delta m\left(\tau\right)\right].\label{eq:m0z_Deltam}
\end{eqnarray}
This expression shows that $m_{z}\left(\tau\right)$ and $\Delta m\left(\tau\right)$
have the same frequency, \textcolor{black}{as confirmed by the green
dots in the inset of Fig.}\textcolor{red}{{} }\ref{fig:Fig_m0z_Dm}. 

Therefore, this simplified model emphasizes the appearance of two
relevant frequencies: the low-frequency of the ordinary precession
and the higher frequency of nutation related with spin fluctuations
at the atomic level driven by the exchange coupling. These two frequencies
clearly show up in Fig. \textcolor{red}{\ref{fig:MSP_Neel}} (blue
wiggles in $m_{z}$). Furthermore, in Eq. (\ref{eq:m0_UA}) we also
see that the spin fluctuations $\bm{\psi}_{i}$ are directly coupled
to the anisotropy parameters $k_{i}$, and this implies that the nutation's
magnitude is not only related to the ratio of surface-to-core spin
number, but also to the value of the anisotropy constants. Note, however,
that the connection between Eq. (\ref{eq:m0_UA}) and Eq. (\ref{eq:LLE-EOSP})
is not a direct one, and one has to eliminate the fast variables $\bm{\psi}_{i}$\emph{,
e.g.} by integration or by making use of their equations of motion
in a perturbative way.

Finally, we have systematically varied the physical parameters ($J_{ij}$,
$k_{i}$) and studied the effect on nutation and the frequencies $f_{p}$,
$f_{c}$ and $f_{n}$. First, we confirm that in the absence of surface
anisotropy (\emph{e.g.} the same uniaxial anisotropy $k_{c}$ for
all spins), no nutation has been observed. This is a direct consequence
of the fact that, in this specific case, there is no magnetic inhomogeneity
in the particle that can lead to a nonuniform effective field. Second,
we find that the precession frequency $f_{p}$ mainly depends on $k_{c}$
since all spins are parallel to each other forming a macrospin that
precesses in the effective uniform field. In general, this would also
include the shape anisotropy and the DC magnetic field. On the other
hand, the frequency $f_{n}$ strongly depends on the exchange coupling
as can be seen in Table \ref{tab:Exchange_Table}. 

\begin{table}
\begin{centering}
\begin{tabular}{cccc}
\hline 
\multirow{2}{*}{$k_{c}$} & \multirow{2}{*}{$k_{s}$} & Precession frequency  & Nutation frequency \tabularnewline
 &  & $f_{p}\left({\rm GHz}\right)$  & $f_{n}\left({\rm THz}\right)$\tabularnewline
\hline 
0.001 & 0.001 & 3.2 & 0\tabularnewline
0.001 & 0.01 & 19 & 1\tabularnewline
0.001 & 0.05 & 86 & 1\tabularnewline
0.001 & 0.1 & 170 & 1\tabularnewline
0.005 & 0.01 & 25 & 1\tabularnewline
0.005 & 0.05 & 93 & 1\tabularnewline
0.005 & 0.1 & 180 & 1\tabularnewline
0.01 & 0.1 & 185 & 1\tabularnewline
\hline 
\end{tabular}
\par\end{centering}
\[
\quad
\]

\begin{centering}
\begin{tabular}{cccc}
\hline 
\multirow{2}{*}{$J_{cs}$} & \multirow{2}{*}{$J_{s}$} & Precession frequency  & Nutation frequency \tabularnewline
 &  & $f_{p}\left({\rm GHz}\right)$  & $f_{n}\left({\rm THz}\right)$\tabularnewline
\hline 
2 & 2 & 25 & 1.5\tabularnewline
1 & 2 & 25 & 1.25\tabularnewline
1 & 1 & 25 & 1\tabularnewline
1 & 0.5 & 25 & 0.75\tabularnewline
1 & 0.1 & 25 & 0.25\tabularnewline
0.5 & 0.5 & 25 & 0.7\tabularnewline
\hline 
\end{tabular}
\par\end{centering}
\caption{\label{tab:Exchange_Table}Precession and nutation frequencies for
fixed values of the exchange couplings $J=J_{cs}=J_{s}=1$ (top) and
for fixed values of core and surface anisotropies $k_{c}=0.005$ and
$k_{s}=0.01$ (bottom).}
\end{table}

We have also performed these calculations for a spherical nanomagnet
which has a different distribution of coordination numbers than in
a cube. The results are qualitatively the same but the nutation frequency
$f_{n}$ is higher. 

\subsection{\label{subsec:IOSP}Comparison with the macroscopic approach to magnetization
nutation}

As discussed in the introduction, magnetization nutation has been
studied by many authors within the macroscopic approach based on Eq.
(\ref{eq:LLGEqt}) augmented by an inertial term proportional to the
second time derivative of the (macroscopic) magnetic moment $\bm{m}$:

\begin{equation}
\frac{d\bm{m}}{d\tau}=\bm{m}\times\left[\bm{h}_{{\rm eff}}-\alpha\,\bm{m}\times\bm{h}_{{\rm eff}}-\frac{\beta}{\tau_{s}}\frac{d^{2}\bm{m}}{d\tau^{2}}\right],\label{eq:LLE-IOSP}
\end{equation}
where the coefficient $\beta$ is often taken proportional to the
damping parameter $\alpha$ and to a phenomenological relaxation time
$\tau_{1}$ related with, \emph{e.g.} the dynamics of the angular
momentum, which is on the order of a femtosecond. In Ref. \onlinecite{mondaletal17prb},
it was shown that the inertial damping results from high-order contributions
to the spin-orbit coupling effect and is related to the Gilbert damping
through the magnetic susceptibility tensor. In the sequel, we shall
use the notation $\tilde{\beta}\equiv\beta/\tau_{s}$ and this macroscopic
model, with the equation of motion (\ref{eq:LLE-IOSP}) and phenomenological
parameter $\tilde{\beta}$, will be referred to as the \emph{inertial
one-spin problem} (IOSP). 

Solving the equation above, in the presence of a DC and AC magnetic
fields, \emph{i.e.} $\bm{h}_{{\rm eff}}=\bm{h}_{DC}+\bm{h}_{AC}$,
Olive \emph{et al. }\cite{Olive_JAP_2012} observed two resonance
peaks, the first of which corresponds to the ordinary large-amplitude
precession at frequency $f_{p}$ and a second resonance peak, at a
much higher frequency $f_{n}$ with smaller amplitude, that was attributed
to the nutation dynamics. A number of other authors made similar observations
by also investigating the IOSP model \cite{PhysRevB.83.020410,PhysRevB.84.172403,PhysRevB.86.020404,PhysRevB.92.140413}.
In Ref. \cite{Olive_JAP_2012} it was suggested that $\omega_{{\rm nutation}}=2\pi f_{n}=1/\beta$.

Let us summarize the situation. On one hand, we have the EOSP model
(applicable when surface effects are not too strong) in which the
dynamics of the net magnetic moment is described by the equation of
motion (\ref{eq:LLE-EOSP}). The solution to the latter only exhibits
two resonance peaks with frequencies $f_{p}$ and $f_{c}$. On the
other hand, we have the IOSP model where the equation of motion is
given by (\ref{eq:LLE-IOSP}) (with the phenomenological parameter
$\tilde{\beta}$) whose solution only provides the two resonance peaks
with frequencies $f_{p}$ and $f_{n}$. Now, the MSP approach, when
treated in its full generality, provides us with a self-consistent
scheme in which all three frequencies appear in a natural manner.
In particular, it shows how nutation with the high-frequency $f_{n}$
sets in, in the presence of surface effects which induce non-collinear
spin configurations and generate high-frequency and small-amplitude
spin-wave excitations. See, for example, a thorough study of spin-wave
excitations in a nanocube in Ref. \onlinecite{BastardisEtal-jpcm2017}.
However, within the MSP approach, the derivation of the equation of
motion for the net magnetic moment $\bm{m}$ (and the spin-wave vectors
$\bm{\psi}_{i}$) is too cumbersome, if not intractable. This issue
will be investigated in the future. Nevertheless, in the case of a
spherical nanomagnet, a Helmholtz equation was derived for the vectors
$\bm{\psi}_{i}$ in Ref. \onlinecite{kachkachi07j3m}, see Eq. (8)
therein, which is nothing else than the propagation equation for the
spin waves described by $\bm{\psi}_{i}$. Now, using the expansion
$\bm{s}_{i}\simeq\bm{m}+\bm{\psi}_{i}$, we may infer that the exchange
contribution $J\Delta\bm{s}_{i}$ is proportional to the second time
derivative of $\bm{m}$ and, as such, the coefficient $\beta\propto1/J$
and thereby $\omega_{{\rm nutation}}\propto J$. The exact relation
will be investigated in a future work.

\begin{figure*}
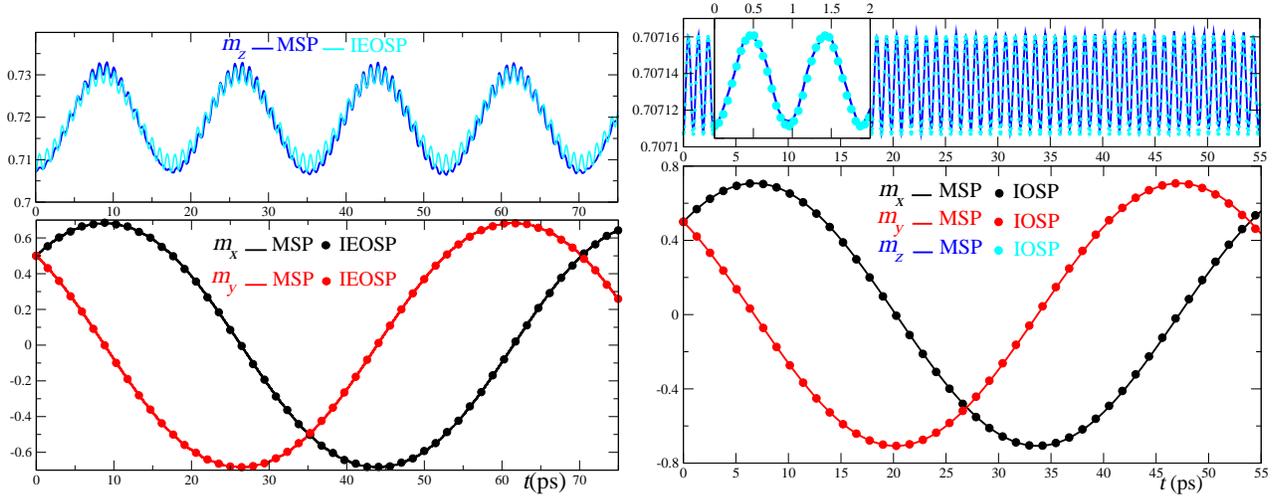

\begin{centering}
\includegraphics[width=0.95\columnwidth]{Figure4a} \includegraphics[width=0.98\columnwidth]{Figure4b} 
\par\end{centering}
\caption{\label{fig:IOSP_NSA}Time evolution of the components of the macroscopic
magnetic moment $\bm{m}$ (dots) and the net magnetic moment (lines)
for MSP. On the left, for Néel surface anisotropy, the MSP results
are compared to the IEOSP model (\ref{eq:LLE-IEOSP}) and on the right,
for uniaxial anisotropy, they are compared to the IOSP model (\ref{eq:LLE-IOSP}).
The inset shows a magnification of the $m_{z}\left(t\right)$ component
with a typical period $\sim0.9\ {\rm ps}$ ($\omega_{{\rm nutation}}\simeq7\ {\rm THz}$).}
\end{figure*}

Nevertheless, there is a specific situation in which we can establish
a clear connection between the MSP approach and the IOSP model. This
is the case of weak surface effects, or equivalently, a quasi-collinear
spin configuration. Indeed, under this condition, we may combine the
EOSP and IOSP models and write an equation of motion whose solution
renders all three frequencies, $f_{p},f_{c}$ and $f_{n}$. More precisely,
we start from Eq. (\ref{eq:LLE-EOSP}) with the effective field $\bm{h}_{{\rm eff}}=2k_{2}m_{z}\bm{e}_{z}-4k_{4}\left(m_{z}^{3}\bm{e}_{z}+m_{y}^{3}\bm{e}_{y}+m_{x}^{3}\bm{e}_{x}\right)$
and add a term similar to that in Eq. (\ref{eq:LLE-IOSP}) with coefficient
$\tilde{\beta}$, leading to the following equation of motion 
\begin{align}
\frac{d\bm{m}}{d\tau} & =\bm{m}\times\left[2k_{2}m_{z}\bm{e}_{z}-4k_{4}\left(m_{z}^{3}\bm{e}_{z}+m_{y}^{3}\bm{e}_{y}+m_{x}^{3}\bm{e}_{x}\right)\right]\label{eq:LLE-IEOSP}\\
 & \qquad-\tilde{\beta}\bm{m}\times\frac{d^{2}\bm{m}}{d\tau^{2}}.\nonumber 
\end{align}
where again we have $k_{2}=k_{c}\mathcal{N}_{c}/\mathcal{N}$ and
for a cube $k_{4}=\left(1-0.7/\mathcal{N}^{1/3}\right)^{4}k_{s}^{2}/zJ$,
and $\tilde{\beta}=\beta/\tau_{s}$. Henceforth, this model will be
referred to as the \emph{inertial effective one-spin problem} (IEOSP).

Compared with Eq. (\ref{eq:LLE-IOSP}), the field $\bm{h}_{{\rm eff}}$
has been replaced in Eq. (\ref{eq:LLE-IEOSP}) by the effective field
produced by the combined uniaxial and cubic anisotropies, induced
by relatively weak surface effects. Of course, we could also include
an external magnetic field and a demagnetizing field in the EOSP equation.
The advantage of the IEOSP model is twofold: i) it renders the three
resonance peaks at the frequencies $f_{p},f_{c}$ and $f_{n}$ and
ii) it allows us to establish a clear connection between the phenomenological
parameter $\tilde{\beta}$ and the atomistic physical parameters of
the MSP approach, such as the surface anisotropy constant $k_{s}$.

For solving Eq. (\ref{eq:LLE-IEOSP}) one needs to set the initial
velocity for $\bm{m}$. For Néel's anisotropy, the system exhibits
several different velocities, depending on the spin position in the
structure (edge, corner, face or core). In this case, one would have
to setup a global constraint by imposing an initial velocity for the
net magnetic moment (\ref{eq:Macrospin}). In practice, we have found
it sufficient to use the average velocity $\dot{\bm{m}}\left(t=0\right)=\sum_{i}\dot{\bm{s}}_{i}\left(t=0\right)/\mathcal{N}$.
The solution of Eq. (\ref{eq:LLE-IEOSP}) is plotted (in dots) in
Fig. \ref{fig:IOSP_NSA} (left).

In Fig. \ref{fig:IOSP_NSA} we show the results from the MSP, IOSP
and IEOSP models. The parameters for the MSP calculations are the
same as in Fig. \ref{fig:MSP_Neel}, \emph{i.e.} $k_{c}=0.001,$ $k_{s}=0.01$.
On the left, we compare the MSP approach to the IEOSP model Eq. (\ref{eq:LLE-IEOSP})
with $k_{2}=0.00475$, $k_{4}=0.0011$, $\tilde{\beta}=2.25$. On
the right, the MSP approach is compared to the IOSP model (\ref{eq:LLE-IOSP})
with the effective field given in Eq. (\ref{eq:effective_parameters-1})
and parameters $k_{{\rm eff}}=0.00576$, $\tilde{\beta}=2.2$. Note
that instead of using the expression for $k_{{\rm eff}}$ in Eq. (\ref{eq:effective_parameters-1})
one might perform a fitting to the MSP curves. Doing so, we find a
slight discrepancy in $k_{{\rm eff}}$ (here $0.00585$) as well as
in the initial velocities $\dot{m}_{\alpha}\left(t=0\right),\alpha=x,y,z$.
This is most likely due to the fact that the velocity average does
not exactly account for the spin non-collinearities. All in all, the
results from the MSP approach are in very good agreement with those
rendered by the macroscopic model, either IOSP or IEOSP, upon using
the corresponding effective field for the given anisotropy configuration
in MSP, namely (\ref{eq:HamUA-UA}) or (\ref{eq:HamUA-NSA}), respectively.
In Fig. \ref{fig:IOSP_NSA} (left), the MSP approach with the anisotropy
model (\ref{eq:HamUA-NSA}) is in good agreement with the IEOSP model
with a given parameter $\tilde{\beta}$. Both models exhibit the three
frequencies $f_{p},f_{c}$ and $f_{n}$. Regarding the nutation with
frequency $f_{n}$, there is a slight discrepancy in amplitude between
the two models. As mentioned above, this is attributed to the average
over the initial velocities. In Fig. \ref{fig:IOSP_NSA} (right) we
see that, for MSP with the anisotropy model (\ref{eq:HamUA-UA}),
the IOSP model (\ref{eq:LLE-IOSP}) with the effective field (\ref{eq:effective_parameters-1}),
recovers the two resonance peaks with $f_{p}$ and $f_{n}$. We draw
attention of the reader to the difference in time scale and amplitude
for the $z$ component. Indeed, the oscillations of the $z$ component
on the right are to be identified with the wiggles of the same component
on the left panel. In Ref. \cite{Olive_JAP_2012} the authors argued
that $\omega_{{\rm nutation}}=1/\beta$. Here, from Fig. \ref{fig:IOSP_NSA}
(right) we extract $\beta\simeq1.43\times10^{-13}{\rm s}$ which should
be compared to $\tilde{\beta}\tau_{s}\simeq1.5\times10^{-13}{\rm s}$,
showing a good agreement. 

Finally, the major difference between the results on the left and
right panels is related with the frequency $f_{c}$. This implies
that the model with uniaxial anisotropy, same easy axis but different
constants in the core and the surface, cannot account for this frequency.
This confirms the fact that the latter is related with the inhomogeneity
of the on-site anisotropy easy direction and thereby with the cubic
effective anisotropy as a first correction to surface effects.

\begin{figure*}
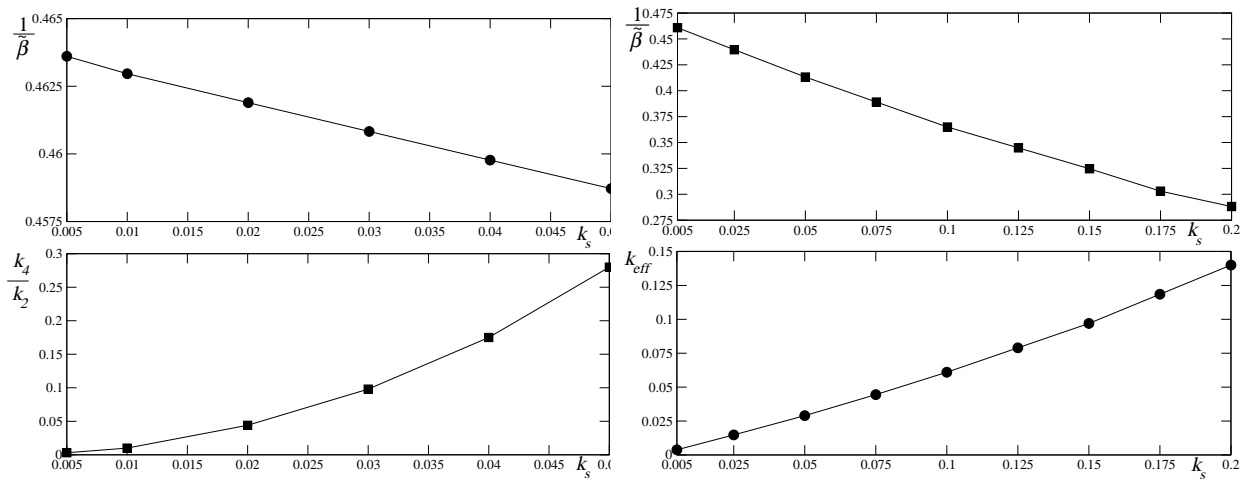

\begin{centering}
\includegraphics[width=0.95\columnwidth]{Figure5a}\includegraphics[width=0.95\columnwidth]{Figure5b}
\par\end{centering}
\caption{\label{fig:Beta_UA}Left: $k_{4}/k_{2}$ and $1/\tilde{\beta}$ against
$k_{s}$. Right: $k_{{\rm eff}}$ and $1/\tilde{\beta}$ against $k_{s}$. }
\end{figure*}

In general, the relation between $\tilde{\beta}$ and the frequency
$f_{n}$, within the MSP approach, is difficult to derive analytically
since $\tilde{\beta}$ depends on the atomic parameters. Nevertheless,
we have tried to establish a quantitative correspondence between the
phenomenological parameter $\tilde{\beta}$ and the microscopic parameters
such as $k_{s},k_{c}$ or the effective parameters $k_{2},k_{4}$
that appear in Eq. (\ref{eq:LLE-EOSP}). Accordingly, in Fig. \ref{fig:Beta_UA}
we plot $1/\tilde{\beta}$ as the result of the best fit between the
MSP and IEOSP models. On the right panel of Fig. \ref{fig:Beta_UA},
this is done for the uniform uniaxial anisotropy model (\ref{eq:HamUA-UA})
and on the left panel for the anisotropy model in Eq. (\ref{eq:HamUA-NSA}).
These results show that $1/\tilde{\beta}$ is nearly linear in $k_{s}$
and that the value of the phenomenological parameter $\tilde{\beta}$
involved in the IEOSP model can be estimated for a given value of
the surface anisotropy constant $k_{s}$, which is an input parameter
of the MSP approach.

\begin{figure}[t]
\begin{centering}
\includegraphics[scale=0.3]{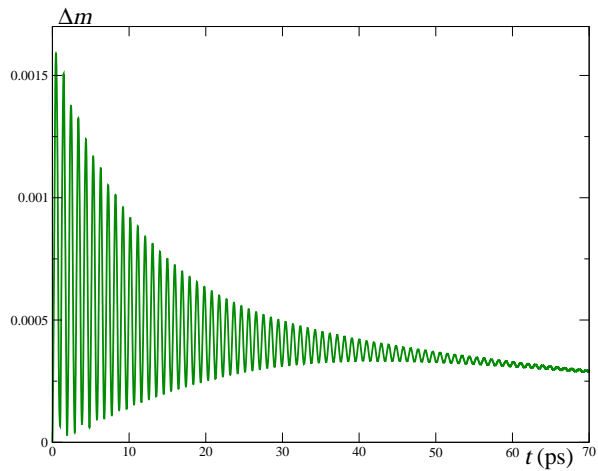} 
\par\end{centering}
\caption{\label{fig:Damping_Cube}Time evolution of the magnetization deficit,
showing the damping effect. Same parameters as in Fig. \ref{fig:MSP_Neel}.}
\end{figure}

Finally, we have investigated the effect of damping with parameter
$\alpha$ {[}see Eq. (\ref{eq:LLE-MSP}){]} within the MSP approach.
The results are shown in Fig. \ref{fig:Damping_Cube} for the magnetization
deficit. Together with the $3D$ picture in Fig. \ref{fig:DampedNutation},
this indicates how the spin fluctuations and thereby $\Delta m$ decays
in time towards zero. 
 This result is obviously in agreement with those of Fig. 1(b) in
Ref. \onlinecite{oliveetal12apl}. We would like to emphasize, though,
that the IOSP approach in its actual formulation cannot account for
the magnetization nutation in the absence of damping because the coefficient
$\beta$ appearing in Eq. (\ref{eq:LLE-IOSP}) before the inertial
term $d^{2}\bm{m}/dt^{2}$ is proportional to damping and thus vanishes
when the latter does. This is one of the major discrepancies with
the MSP approach since the latter does produce magnetization nutation
even in the absence of such a damping ($\alpha=0$). However, within
the MSP approach the surface-induced nutation is due to local spin
fluctuations and is thus affected by the spin-spin correlations or
multi-magnon processes which cause damping effects and relaxation
of the magnetization deficit. But in the absence of a coupling of
the spin subsystem to the lattice, referred to in Ref. \cite{suhl98ieee}
as the direction relaxation, these damping effects are not dealt with
in this work and this is why when we set $\alpha=0$ the time evolution
of $m_{\alpha}$ or $\Delta m$ is undamped, but does exhibit nutation.

\section{\label{sec:Conclusion-and-perspectives}Conclusion and perspectives}

We have proposed an atomistic approach for studying the effects of
surface anisotropy and investigating nutation in the magnetization
dynamics in ferromagnetic nanoparticles. We have then shown that because
of these effects, which induce spin noncolinearities leading to nonuniform
local effective fields, the magnetization dynamics exhibits several
resonance peaks. In addition to the ordinary precessional motion with
frequency $f_{p}\sim10\,{\rm GHz}$, we have shown that the dynamics
of the net magnetic moment exhibits two more resonance peaks with
frequencies $f_{c}$ and $f_{n}$ which are higher than the FMR frequency.
Indeed, $f_{c}=4\times f_{p}\sim40\,{\rm GHz}$ is related with the
oscillations of the particle's magnetic moment between the minima
of the effective potential induced by weak surface anisotropy. On
the other hand, the much higher frequency $f_{n}\sim1\,{\rm THz}$
is attributed to the magnetization fluctuations at the atomic level
driven by exchange coupling which becomes relevant in the presence
of strong nonuniform spin configurations. 

We have compared our results on nutation induced by surface effects
with those rendered by the macroscopic approach based on the Landau-Lifshitz-Gilbert
equation augmented by an inertial term (proportional to the second-order
time derivative of the macroscopic moment) with a phenomenological
coefficient. The good agreement between the two models makes it possible
to estimate this coefficient in terms of the atomistic parameters
such as the surface anisotropy constant. In brief, the atomistic approach
provides a new origin for the magnetization nutations and a global
and a self-consistent picture that renders all three frequencies.

In the case of not-too-strong surface effects, an effective model
renders two frequencies $f_{p}$ and $f_{c}$. On the other hand,
the Landau-Lifshitz-Gilbert equation with an inertial term only renders
the frequencies $f_{p}$ and $f_{n}$. Now, in the case of arbitrary
surface effects, it is a rather difficult task to derive an effective
equation of motion for the magnetization dynamics. As such, we have
proposed an intermediate model that starts from the effective model
established for weak surface effects and added magnetization inertia
through the term proportional to second-order time derivative of the
magnetization. Then, we have shown that this macroscopic model is
in very good agreement with the atomistic approach and renders all
resonance peaks and their frequencies. This establishes a clear quantitative
connection between the phenomenological parameters of the macroscopic
approach to the atomistic energy parameters. 

Our final word is devoted to the possibility of experimental observation
of nutation in magnetization dynamics. First of all, establishing
the fact that surface effects do induce magnetization nutation may
provide us with an additional means for observing the latter. Indeed,
surface effects on ferromagnetic resonance in nanoparticles have been
studied for a few decades now. For example, the authors of Ref. \onlinecite{respaudetal99prb}
reported on high-frequency FMR ($115-345\,{\rm GHz}$) spectra for
ultra-fine cobalt particles and inferred rather small values of the
transverse relaxation time $\tau_{\perp}$ which suggests that this
should be due to an inhomogeneous precession caused by (relatively
weak) surface spin disorder. There are several other publications
on FMR measurements on magnetic nanoparticles \cite{gazetal98j3m,shilovetal99prb,schsch07jncs,Schoeppneretal14jap,ollefsetal15jap,poprai16prb}.
However, these measurements can only capture the two frequencies $f_{p}$
and $f_{c}$. Nevertheless, the observation of the frequency $f_{c}$,
which is on the order of tens of GHz, should be an easy matter using
a network analyzer with variable frequency covering this range. Doing
so would clearly prove the existence of the first nutation motion
induced by spin disorder as a consequence of surface anisotropy. A
variant of the FMR spectroscopy, called Magnetic Resonance Force Microscopy
\cite{Sidles_RevModPhys,pigeauetal12prl,lavenantetal2014nanofab},
yields a highly sensitive local probe of the magnetization dynamics
and consists in mechanically detecting the change in the longitudinal
fluctuations of the magnetization, \emph{i.e.} $\Delta m_{z}$. This
would be particularly suited for detecting the fluctuations in $m_{z}$
seen in Figs. \ref{fig:MSP_Neel} and \ref{fig:IOSP_NSA}, if not
for the mismatch in the frequency range. Now, the frequency $f_{n}$
is rather in the optical range and we wonder whether the corresponding
oscillations could be detected by coupling the magnetization of the
nanoparticle to a plasmonic nanoparticle of gold or silver, thus making
use of the magneto-plasmonic coupling evidenced in many hybrid nanostructures
\cite{gonzalezetal08small,temnovetal10natphot,armellesetal13aom}.
Graphene plasmons is another promising route for detection of THz
radiation \cite{Bandurin_2018arXiv}.
\begin{acknowledgments}
We would like to acknowledge useful discussions with J.-E. Wegrowe
on their early work on magnetization nutation. 
\end{acknowledgments}

\bibliography{hkbib}

\end{document}